\documentclass[amssymb,amsmath,groupedaddress,pra, 12pt]{revtex4-2}

\usepackage[english]{babel}
\usepackage{graphicx}
\usepackage{color}
\usepackage{hyperref}
\usepackage{bm}
\usepackage{amsmath}
\usepackage{amssymb}
\usepackage{url}
\usepackage{natbib}

\usepackage[T1,T2A]{fontenc}

\DeclareGraphicsExtensions{.eps,.pdf,.png,.jpg}

\begin{document}
	
	\title{The Brillouin flow in a smooth-bore magnetron fed by split cathode }
	
	\author{Y. Bliokh}
	\email{bliokh@physics.technion.ac.il}
	\affiliation{Physics Department, Technion, Israel Institute of Technology, Haifa 320003, Israel}

	\author{J. G. Leopold}
	\affiliation{Physics Department, Technion, Israel Institute of Technology, Haifa 320003, Israel}
	
	\author{Ya. E. Krasik}
	\affiliation{Physics Department, Technion, Israel Institute of Technology, Haifa 320003, Israel}

\begin{abstract}

	Explosive emission from an axial cathode of a relativistic magnetron produces  plasma, the radial expansion of which can cause pulse shortening. In a split cathode fed magnetron, the electron source and its explosive plasma are outside the space where the high power microwave producing interaction occurs. This electron source is a longitudinal annular electron column expanding radially. This expansion simulates the radial emission from an axial cathode. A mathematical model and numerical simulations are presented which enable to calculate the parameters of this electron column, its density, angular velocity, and potential distributions. The  Hull Cutoff and Buneman-Hartree modified conditions applicable to the split cathode magnetron are formulated.

\end{abstract}	
	
\maketitle

\section{Introduction}

The magnetron is the most studied device where high-frequency electromagnetic oscillations are excited by the flow of electrons moving in crossed electric and magnetic fields. In conventional non-relativistic magnetrons,  electrons are emitted by a cylindrical thermionic cathode located coaxially with an anode. A radial electric field between the anode and cathode accelerates the electrons toward the anode, while an external axial magnetic field deflects the electrons in an azimuthal direction. If the magnetic field exceeds some critical value $H_{\rm HC}$, which depends on the applied anode-cathode potential $U_0$, the emitted electrons do not reach the anode. The dependence $H_{\rm HC}(U_0)$ is defined by the cathode and anode radii and is known as the \textit{Hull Cutoff} (HC) criterion \cite{SC-Hull-1921,  Slater-1946, Slater-1963, SC-Buneman-1961, SC-Lau-2010}.  

When the HC criterion is fulfilled, the electrons form around the cathode a rotating cloud, whose outer boundary does not reach the anode surface. Two different theoretical models of the electron motion in the cloud exist \cite{ Brillouin-1951}. In a ``double-stream’’ model, there is an electron flow from the cathode and an equal flow in the opposite direction. In a ”single stream’’  model, a laminar electron flow rotates around the cathode. This is the so-called \textit{Brillouin flow}. Numerous numerical simulations and theoretical studies indicate that the Brillouin flow is realized \cite{Slater-1946, SC-Buneman-1961, Orzechowski-1979, Christenson-1996, Christenson-1994}.

The rotating electron flow excites effectively the electromagnetic eigenmode of the cavity magnetron, when the electron azimuthal velocity $v_\theta$ at the anode surface is equal to the azimuthal phase velocity, $v_{ph}$, of the mode. The condition $v_\theta=v_{ph}$ is known as the \textit{Buneman-Hartree} (BH) condition\cite{SC-Buneman-1961,Lovelace-1985,Benford-2007}. 

The \textit{relativistic} magnetron is a high-current, high-voltage version of the conventional magnetron, which is used as a source of microwave oscillations of hundreds MW and GW power levels\cite{Benford-2007}. The high voltage is essential for the generation of explosive emission plasma at the cathode surface. This plasma serves as a source of electrons and provides a high density of the emitted current. However, this plasma has disadvantages. The electrons are emitted and accelerated towards the anode from the boundary of this dense, conducting plasma, rather than from the cathode surface. The radial expansion of the plasma makes the effective cathode radius a time-varying quantity which leads to violation of the HC and BH conditions. As a result, the microwave pulse is shorter than the duration of the applied voltage pulse. This is the so-called \textit{pulse shortening} effect\cite{SC-Price-1998,SC-Price-1998a}. To extend the microwave pulse, it is necessary to eliminate the effect of the explosion emission plasma inside the anode interaction space, and the split cathode is a possible solution.

In this article the smooth-bore magnetron is considered because the slow-wave structure at the anode (cavities) does not affect the electron motion in the Brillouin flow\cite{Slater-1963, Palevsky-1979}. 

Despite significant progress in experimental investigations\cite{SC-Liziakin-2024,SC-Bliokh-2022,SC-Krasik-2022}, numerical simulati\-ons\cite{SC-Leopold-2023}, and theoretical studies\cite{SC-Leopold-2020a}, achieved during the last decade, the formation of the electron flow in a SC fed magnetron, even in a smooth-bore magnetron, has not yet been studied. However, the structure of the laminar electron flow (the Brillouin flow) determines the BH and HC conditions, crucial for crossed-field electronic devices. 

\begin{figure}[tbh]
	\centering \scalebox{0.5}{\includegraphics{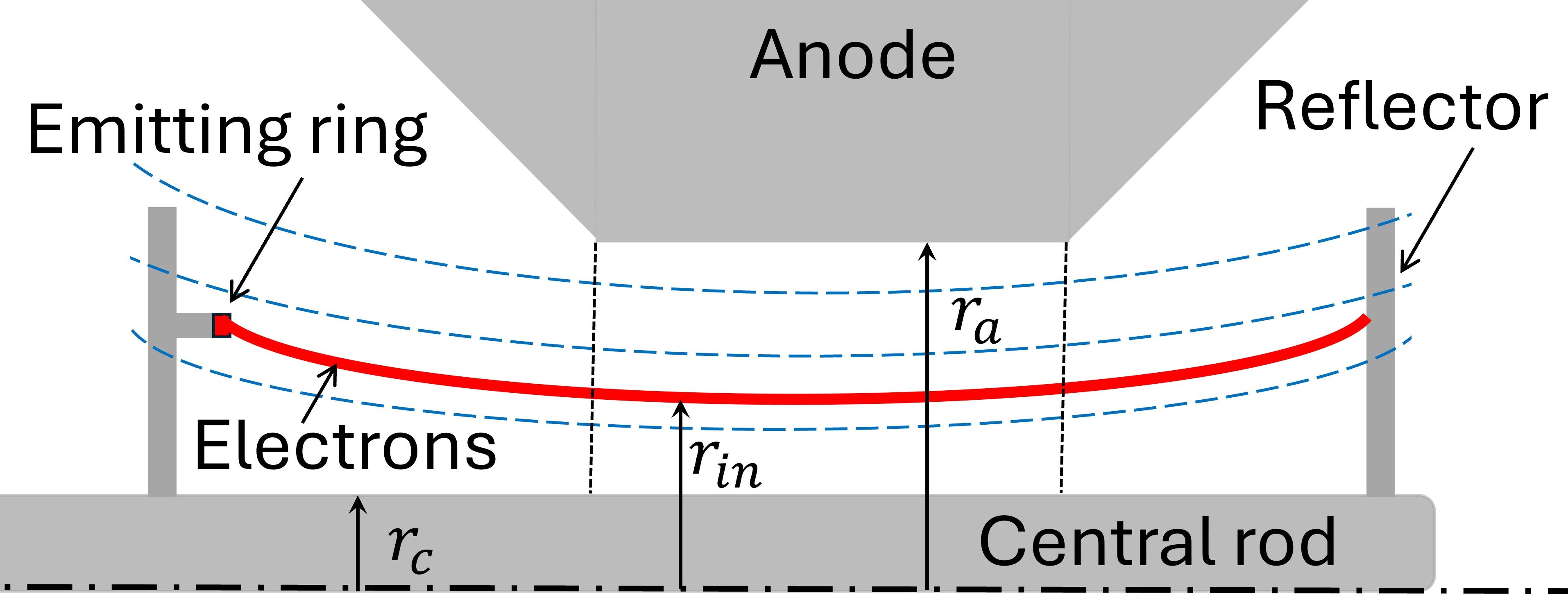}}
	\caption{A coaxial diode or else a smooth bore magnetron with a split cathode. The annular electron beam, emitted from the emitting ring, is magnetized by an external axial magnetic field $H_0$ (shown by blue dashed lines). The electrodes and  magnetic field are configured so that in the interaction space (space between the dotted lines), the magnetic field is homogeneous, and electrons form a laminar Brillouin flow.  }
	\label{fig1}
\end{figure}

Schematically a SC is shown in Fig.~\ref{fig1}. The split cathode consists of an annular emitting ring  placed at a distance outside the upstream end of an anode and a reflector placed at a distance from the downstream end of the anode, which suppresses axial electron flow further downstream. These two parts are connected by a central conducting rod so that the emitting ring, reflector, and rod have the same potential. The SC is placed coaxially with the anode. The emitted annular electron beam is magnetized by an external axial magnetic field $H_0$. It is suggested, that the electrodes and the magnetic field are configured so that in the anode region, where magnetic field is homogeneous, electrons form a laminar Brillouin flow \cite{SC-Brillouin-1941}, just as it takes place in a magnetron gun \cite{SC-Waters-1963}. The emitted electrons are reflected from the reflector and return to the cathode, forming an annular electron column with two counter-propagating electron fluxes. The formation and properties of the electron column in the longitudinally homogeneous  interaction space are the subject of the present studies.

\section{Physical model \label{Physical model}}

It is assumed below that the electric and magnetic fields, and space charge of the electrons in the system are axially symmetric. What happens in the interaction space when some potential $U_0$ is applied to the anode? The first group of emitted electrons propagates along the entire system up to the reflector from where they turn back. By then the  space is already filled with electrons the space charge of which changes the potential and the electric field along the propagating path of the reflected electrons. Because of this, the trajectories of the electrons moving toward the emitting ring are shifted closer to the anode. This process of displacing earlier emitted electrons toward the anode by the space charge of later emitted ones occurs continuously.

It is assumed that all emitted electrons appear in the  interaction space at the same radius $r_{\rm in}$. However, the potential $\varphi$ at this radius, $\varphi(r_{\rm in})$, depends on the space charge already accumulated there. It means that the initial kinetic energy of any electron, which oscillates between the cathode and reflector, is defined by its injection time. This is the first main peculiarity, which makes the electron dynamics in a SC fed magnetron different  from that with a radially emitting cathode.  Indeed, in a smooth bore coaxial diode or a magnetron fed by an axial cathode all electrons in the  interaction space have the same zero initial kinetic energy.   

Because of the axial symmetry of the system, the canonical angular momentum of an electron is a conserved quantity, the value of which is determined by its angular velocity at time $t_0$ when the particle appears in the interaction space. In the Brillouin flow this velocity depends on the value of the radial electric field $E_r$ at the injection point $r_{\rm in}$. As mentioned, the value of the electric field $E_r(r_{\rm in})$ depends on the charge, accumulated earlier. In other words, each electron is characterized by its own canonical angular momentum. This is an additional important distinction from a  coaxial cathode, which emits electrons in the radial direction with the same zero angular momentum.

Thus, in a smooth bore magnetron (coaxial diode) the total energy and the canonical angular momentum of the electrons depends on their injection time. This makes determining the resulting steady-state radial distribution of the potential, electron density, and electron velocity a much more complicated problem.  The proposed mathematical model and the scheme of the numerical simulations allows one to describe a transient process for the formation of a Brillouin flow and determine  parameters of the resulting steady state such as the radial distribution of the electron density, the angular velocity, etc.

\section{Mathematical model \label{Math model}}       

The  mathematical model is based on the following assumptions.
\begin{enumerate}
	\item {Only an axially symmetric configuration is considered.}
	\item {The electron motion is non-relativistic. This restriction is not fundamental, it} simplifies calculations and allows one to separate the longitudinal from the transverse electron motion.	
	\item {The electron motion and potential distribution are considered in the anode space, where the system is longitudinally homogeneous.}
	\item {Electron motion is considered only in the transversal plane.} 	
	\item The width of the emitting ring of SC is small compared with the rod-anode gap, so that all electrons appear at $t_0$ in the interaction space at the same radius $r_{\rm in}$ (thereafter the injection radius). 
	\item {The electrons enter the interaction space with zero radial velocity. The angular velocity matches the local radial electric field, i.e., corresponds to laminar Brillouin flow. }
	\item {The characteristic time of the charge accumulation in the interaction space is large compared to the electron traveling time between the anode and the reflector. } 
\end{enumerate}  

At first glance, the number of restrictions listed above shrinks the applicability range of the theory. However, under close examination, all these assumptions are widely used in theoretical studies of the stationary Brillouin flow in magnetrons or coaxial diodes. Only assumptions~\#6 and \#7 relate to the non-stationary transit  process, which is not usually considered.

Let us represent the electron column as a set of discrete charged cylindrical layers  of zero thickness and surface charge $-\sigma_i$ ($\sigma>0$, the electron charge is $-e$). This charged layer of radius $r_i$, situated in the space between two conducting cylindrical surfaces, the central rod and the anode, at equal potentials, produces a radial electric field 
\begin{eqnarray}\label{eq1}
	E_i^{(+)}(r)=-4\pi\sigma_i (r_i/r)\ln(r_i/r_c)/\ln(r_a/r_c),\hspace{3mm}r>r_i\nonumber\\
	E_i^{(-)}(r)=-4\pi\sigma_i (r_i/r)\ln(r_i/r_a)/\ln(r_a/r_c),\hspace{3mm}r<r_i.
\end{eqnarray}
where $r_c$ and $r_a$ are the central rod and anode radii, respectively, and $E_i^{(+)}(r)$ and $E_i^{(-)}(r)$ are the electric fields above and below the electron charge layer. In addition to the space charge fields (\ref{eq1}), an external electric field  $E_{ext}(r)=-U_0/[r\ln(r_a/r_c)]$ exists, where $U_0$ is the anode potential relative to the central rod  which is at zero potential. Thus, the total electric field $E(r)$ is
\begin{equation}
	\label{eq2}
	E(r)=E_{ext}(r)+\sum_{r_i<r}E_i^{(+)}(r)+\sum_{r_i>r}E_i^{(-)}(r). 
\end{equation} 

Electron motion in the crossed radial electric, $E(r)$,  and the longitudinal magnetic,  $H_0$, fields is governed by the following equations:
\begin{eqnarray}
	\frac{d^2r}{dt^2}=-\frac{e}{m}E(r)+\frac{1}{r}v_\varphi^2-\omega_Hv_\varphi,\label{eq3}\\
	rv_\varphi-\frac{1}{2}\omega_Hr^2\equiv P={\rm const}.\label{eq4}
\end{eqnarray} 
Here $v_\varphi$ is the angular velocity,  $\omega_H=eH_0/mc$ is the cyclotron frequency, and $P$ is the canonical angular momentum, which is an integral of motion due to the axial symmetry of the fields. Using Eq.~(\ref{eq4}), one can eliminate the angular velocity and rewrite Eq.~(\ref{eq3}) as follows:
\begin{equation}\label{eq5}
		\frac{d^2r}{dt^2}=-\frac{e}{m}E(r)-\frac{1}{4}\omega_H^2r+\frac{P^2}{r^3}.
\end{equation}	

Following assumption~\#5, any electron appears in the interaction space at  the injection radius $r=r_{\rm in}$ which by assumption~\#6 is the equilibrium radius for which the right-hand-side of Eq.~(\ref{eq5}) is equal to zero. This condition defines the value of the canonical angular momentum $P$ to be:
\begin{equation}\label{eq6}
	P^2=\frac{1}{4}\omega_H^2r_{\rm in}^4+\frac{e}{m}E_{\rm in}r_{\rm in}^3,
\end{equation}	
where $E_{\rm in}=E(r_{\rm in},t_{\rm in})$ and $t_{\rm in}$ is the entry time. Substituting  $P^2$ in Eq.~(\ref{eq5}) its value Eq.~(\ref{eq5}) gives:
\begin{equation}
	\label{eq7}
	\frac{d^2r}{dt^2}=-\frac{e}{m}\left[E(r)-E_{\rm in}\frac{r_{\rm in}^3}{r^3}\right]-\frac{1}{4}\omega_H^2r\left(1-\frac{r_{\rm in}^4}{r^4}\right).
\end{equation}

Any newly injected charged layer increases the electric field everywhere at $r>r_{\rm in}$, so the equilibrium radii of earlier injected electrons increase as well. In such a manner the electron column expands toward the anode.   	
Note that the total charge associated with any layer remains constant, so that the surface charge density $\sigma_i$ in Eq.~(\ref{eq1}) varies with the layer radius as $\sigma_i(r_i)=\sigma_i^{(0)}r_{\rm in}/r_i$, where $\sigma_i^{(0)}=\sigma_i(r_{\rm in},t_{\rm in})$. 

The proposed scheme of numerical simulation is similar to that applied to describe  the transient process of the Brillouin flow formation in a planar magnetron\cite{SC-Buneman-1961}. At the beginning, when the anode space is empty, the electric field at the entry radius $r_{\rm in}$ is determined by the external field $E_{ext}(r_{\rm in})$. The electrons of this first layer are at once placed in an equilibrium orbit.  

The second layer is injected at the same radius $r_{\rm in}$, but the electric field $E(r_{\rm in})$ is already distorted by the presence of the first layer: $E(r_{\rm in})=E_{ext}(r_{\rm in})+E_1^{(-)}(r_1)$, where $r_1$ is the position of the first layer, which was equal to $r_{\rm in}$  at the injection time, and $E_1^{(-)}(r)$ is defined by Eq.~(\ref{eq1}). It will be discussed later how the charge of layers, $q_i=2\pi r_{\rm in}\sigma_i^{(0)}$, should be defined. Further, the two equations of motion  of these two layers Eq.~(\ref{eq7}) are solved simultaneously with the  initial conditions $r_1(0)=r_2(0)=r_{\rm in}$, $v_1(0)=v_2(0)=0$, where $r_2(t)$ is the second layer position, and $v_{1,2}=dr_{1,2}/dt$ are the radial velocities of the layers. The subsequent layers are added in the same manner.

The layer charge $q_i$ is defined by the current emitted by the cathode during the  time chosen to be the interval between the injection of the layers.  This time interval is determined by the number of periods of
electron oscillations between the cathode and reflector until steady state distribution of the
space charge and potential is obtained in this layer in the interaction space. Assuming, that electron emission is space-charge limited, it is reasonable to expect that the emitted current depends eventually on the potential   $\varphi(r_{\rm in})$  at the injection radius at the specific injection time, neglecting processes in the flow formation between interaction space and cathode (see Fig.~\ref{fig1}). Because in this model there are no electrons between  the central rod and the injection radius $r_{\rm in}$, this potential depends only on the value of the electric field $E(r_{\rm in})$:
\begin{equation}
	\label{eq10}
	\varphi(r_{\rm in})=-E(r_{\rm in})r_{\rm in}\ln\left(r_{\rm in}/r_c\right).
\end{equation} 

It will be assumed in Section IV that relation between the emitted current $I$ and the potential $\varphi(r_{\rm in})$ is the same as that defined by the Child-Langmuir law,  $I\propto\varphi^{3/2}(r_{\rm in})$, so that the charge $q_i$ is proportional to $[E_c^{(i)}]^{3/2}$: $q_i=c_q[E_c^{(i)}]^{3/2}$, where $c_q$ is the proportionality coefficient. 

\section{Numerical results \label{Numerical-results}}

First, let us define conditions, when the Brillouin flow can exist. The first injected layer possesses the smallest canonical angular momentum $P$,  defined by Eq.~(\ref{eq6}). Indeed, the electric field $E_{\rm in}$ at the injection radius $r_{\rm in}$ decreases with increasing number of earlier injected layers, that increases the value of the momentum in accordance with Eq.~(\ref{eq6}). For the first layer, the electric field $E_{\rm in}$ is defined only by the external potential, $E_{\rm in}=-U_0/r_{\rm in}\ln(r_a/r_c)$. If the condition
\begin{equation}\label{Brillouin}
	P^2>0
\end{equation}   
is satisfied for the first injected layer, it is also correct for all subsequent layers. The condition Eq.~(\ref{Brillouin}) determines the maximal value, $U_{\rm max}$, of the anode potential $U_0$ such that Brillouin flow can be created:
\begin{equation}\label{Brillouin-2}
	\frac{eU_0}{mc^2}\leq \frac{eU_{\rm max}}{mc^2}=\frac{1}{4}\frac{\omega_H^2r_{\rm in}^2}{c^2}\ln(r_a/r_c).
\end{equation} 

The kinetic energy $w$ of electrons in the injected layer is defined by the potential $\varphi(r_{\rm in})$, $w=e\varphi(r_{\rm in})$. Part of this energy is associated with the longitudinal velocity $w_\parallel=mv_\parallel^2/2$. The rest, $w_\perp$, is associated with the angular velocity $v_\varphi$: $w_\perp=mv_\varphi^2/2$.  The first injected layer possesses the maximal angular velocity
\begin{equation}\label{Velocity}
	v_\varphi=\frac{1}{2}\omega_Hr_{\rm in}\left[1-\sqrt{1-U_0/U_{\rm max}}\right].
\end{equation}
(see Eq.~(\ref{eq3})) and the kinetic energy $w_\perp$ of electrons comprising this layer is maximal. Evidently, $w_\perp\leq w=e\varphi(r_{\rm in})=eU_0\ln(r_{\rm in}/r_c)/\ln(r_a/r_c)$. The condition $w_\perp\leq w$ and Eq.~(\ref{Brillouin-2})
are compatible when $2\ln(r_{\rm in}/r_c)>1$, or
\begin{equation}\label{Brillouin-4}
	r_{\rm in}>r_c\sqrt{e}\simeq 1.65 r_c.
\end{equation}
If the opposite to inequality (\ref{Brillouin-4})  is fulfilled, then the permissible value of the potential is smaller than that defined by Eq.~(\ref{Brillouin-2}). For example, when $r_{\rm in}\simeq 1.3 r_c$, then the maximal permissible value of the potential is about 4\% less that defined by Eq.~(\ref{Brillouin-2}). For simplicity, it is assumed below that the condition Eq.~(\ref{Brillouin-4}) is satisfied.

Accumulation of the charge in the interaction space results in the decrease of the electric field at the internal boundary of the electron column, at $r=r_{\rm in}$, that is, a decrease potential gap from this radius down to the central rod potential $\varphi(r_c)=0$, and a decrease in the emitted current (charge $q_i$ of a new injected layer) down to zero. Such asymptotic steady-state with zero anode-cathode current can be formed if the external radius $r_{\rm max}$ of the electron column does not reach the anode surface $r=r_a$. 

\begin{figure}[tbh]
\centering \scalebox{0.5}{\includegraphics{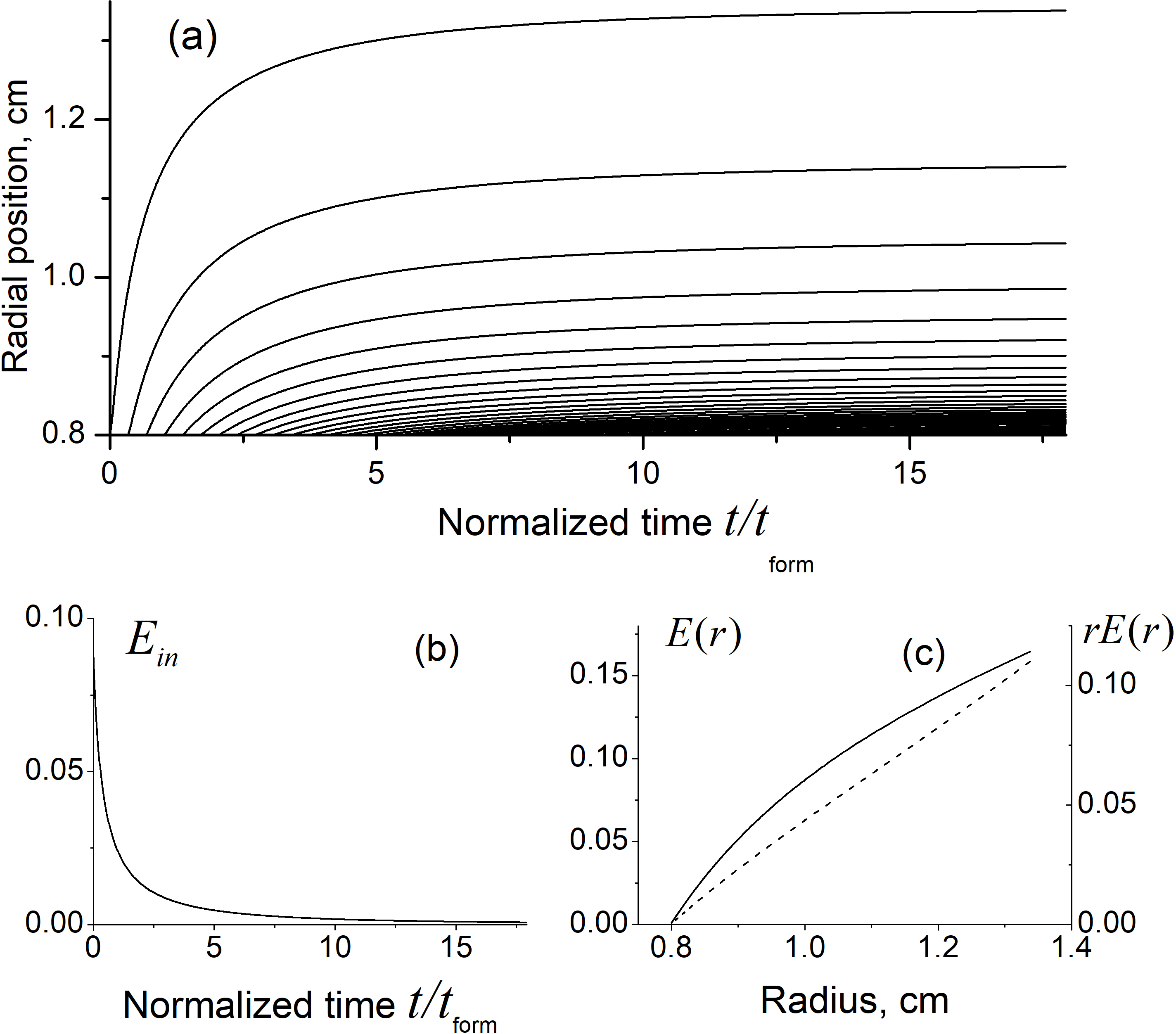}}
\caption{ (a) Trajectories of the layers injected one by one into the anode-rod gap. (b) Evolution of the electric field $E_{\rm in}=E(r_{\rm in})$. (c) Radial distribution of the electric field $E(r)$ across the asymptotic Brillouin flow (number of layers tends to infinity)  (solid line, left axis) and the product $rE(r)$ (dashed line, right axis). The parameters for these calculations are: $r_a=2$~cm, $r_c=0.25$~cm, $r_{\rm in}=0.8$~cm, $H_0=2$~kG, $U_0=200$~kV.    }
\label{fig2}
\end{figure}

The trajectories $r_i(t)$ of the layers, injected one by one into the anode-rod gap, are shown in Fig.~\ref{fig2}a. In Fig.~\ref{fig2}b the temporal evolution of the electric field $E_{\rm in}(t)$ at the inner radius $r_{\rm in}$ of the electron column is drawn, whereas in Fig.~\ref{fig2}c the radial distribution of the electric field for the asymptotic Brillouin state (infinite time) and $rE(r)$ are shown. Note that the characteristic time of the Brillouin flow formation,  $t_{\rm form}$, depends on the specific geometric dimensions of the system, shown schematically in Fig.~\ref{fig1}.

As seen in Fig.~\ref{fig2}a, the outer radius of the electron column expands as the accumulated charge increases, but it does not reach the anode. At the same time the electric field $E_{\rm in}$ approaches zero, i.e., $\varphi(r_{\rm in})\rightarrow \varphi(r_c)$  which corresponds to cathode emission approaching zero. Figure \ref{fig2}c demonstrates an interesting property of the asymptotic state of the electron column, that is, the product $rE(r)$ can be approximated by a linear function of the radius:
\begin{equation}\label{eq12a}
	rE(r)=-\alpha(r-r_{\rm in}).
\end{equation}	
This approximation  determines the electron column's complete radial structure and such parameters as the outer radius $r_m$, electron density, and  angular velocity.
 	 	
Indeed, if the electron density distribution $n(r)$ is known, then one can calculate the distribution of the potential $\varphi(r)$ across the electron column, $r_{\rm in}<r<r_m$. Consider that the potential and field are continuous at the boundary $r=r_m$ and using boundary conditions for the asymptotic steady-state electron flow, $\varphi(r_{\rm in})=\varphi(r_c)=0$ and $\varphi(r_a)=U_0$, the following relation can be derived:
\begin{equation}\label{eq13}
	U_0=I_1(r_m)+I_2\ln(r_a/r_m),
\end{equation} 
where
\[I_1(r)=4\pi e\intop_{r_{\rm in}}^{r_m} \frac{dr^\prime}{r^\prime}\intop_{r_{\rm in}}^{r^\prime}dr^{\prime\prime}\,r^{\prime\prime}n(r^{\prime\prime}),  \]
and
\[I_2=4\pi e\intop_{r_{\rm in}}^{r_m} dr\,rn(r).\]

The outer radius of the electron column, $r_m$, is formed by the first injected layer. As mentioned, for the first layer the electric field $E_{\rm in}$  at the time of injection is defined only by the external potential:
\begin{equation}\label{eq14} 
E_{\rm in}=-U_0/[r_{\rm in}\ln(r_a/r_c)].
\end{equation} 
The electric field $E(r_m)$ is defined by the charge, accumulated in the column:
\begin{equation}\label{eq15}
	E(r_m)=-I_2/r_m.
\end{equation}
Orbits of the electrons comprising this layer are stationary, $v_r=0$, when the right-hand-side of Eq.~(\ref{eq7}) is equal to zero:
\begin{equation*}
	\frac{e}{m}\left[E(r_m)-E_{\rm in}\frac{r_{\rm in}^3}{r_m^3}\right]+\frac{1}{4}\omega_H^2r_m\left(1-\frac{r_{\rm in}^4}{r_m^4}\right)=0,
\end{equation*}
or by considering Eq.~(\ref{eq14}) and (\ref{eq15}),
\begin{equation}\label{eq16}
	\frac{e}{m}\left[I_2-\frac{U_0}{\ln(r_a/r_c)}\frac{r_{\rm in}^2}{r_m^2}\right]=\frac{1}{4}\omega_H^2r_m^2\left(1-\frac{r_{\rm in}^4}{r_m^4}\right).
\end{equation}

The electron density distribution $n(r)$ can be determined using the Poisson equation and the field approximation Eq.~(\ref{eq12a}):
\begin{equation}\label{eq17}
	4\pi en(r)=\alpha/r.
\end{equation}
The integrals $I_1$ and $I_2$ can now be evaluated:
\begin{eqnarray}
I_1=\alpha\left[(r_m-r_{\rm in})-r_{\rm in}\ln(r_m/r_{\rm in})\right],\nonumber\\
I_2=\alpha(r_m-r_{\rm in}). \label{eq18}
\end{eqnarray}
Then, Eq.~(\ref{eq13}) assumes the form:
\begin{equation}\label{eq16a}
	U_0=\alpha r_aD(\rho_m,\rho_{\rm in}),
\end{equation}
where $D(\rho_m,\rho_{\rm in})=\rho_m-\rho_{\rm in}+\rho_{\rm in}\ln\rho_{\rm in}-\rho_m\ln\rho_m$, and $\rho$ is the radial distance, normalized to the anode radius $r_a$: $\rho_m=r_m/r_a$, $\rho_{\rm in}=r_{\rm in}/r_a$. 

With Eqs.~(\ref{eq18}) and (\ref{eq16a}) one can rewrite Eq.~(\ref{eq16}) in the following dimensionless form:
\begin{equation}\label{eq19}
	\frac{\Phi_0}{|\ln\rho_c|}\left[\frac{(\rho_m-\rho_{\rm in})\ln\rho_c}{D(\rho_m,\rho_{\rm in})}+\frac{\rho_{\rm in}^2}{\rho_m^2}\right]+\frac{1}{4}\Omega_H^2\rho_m^2\left(1-\frac{\rho_{\rm in}^4}{\rho_m^4}\right)=0.
\end{equation} 
Here $\Phi_0=eU_0/mc^2$ is the normalized anode potential and $\Omega_H=\omega_Hr_a/c$ is the normalized magnetic field magnitude.
The applicability domain of this equation is defined by inequality (\ref{Brillouin}):
\begin{equation}
	\Phi_0\leq \Phi_{\rm max}=\frac{1}{4}\Omega_H^2\rho_{\rm in}^2|\ln\rho_c|.\label{eq20}
\end{equation}
Equation (\ref{eq19}) determines the outer radius $\rho_m$ of the electron column as function of the applied magnetic field and anode potential. 

\begin{figure}[tbh]
	\centering \scalebox{0.33}{\includegraphics{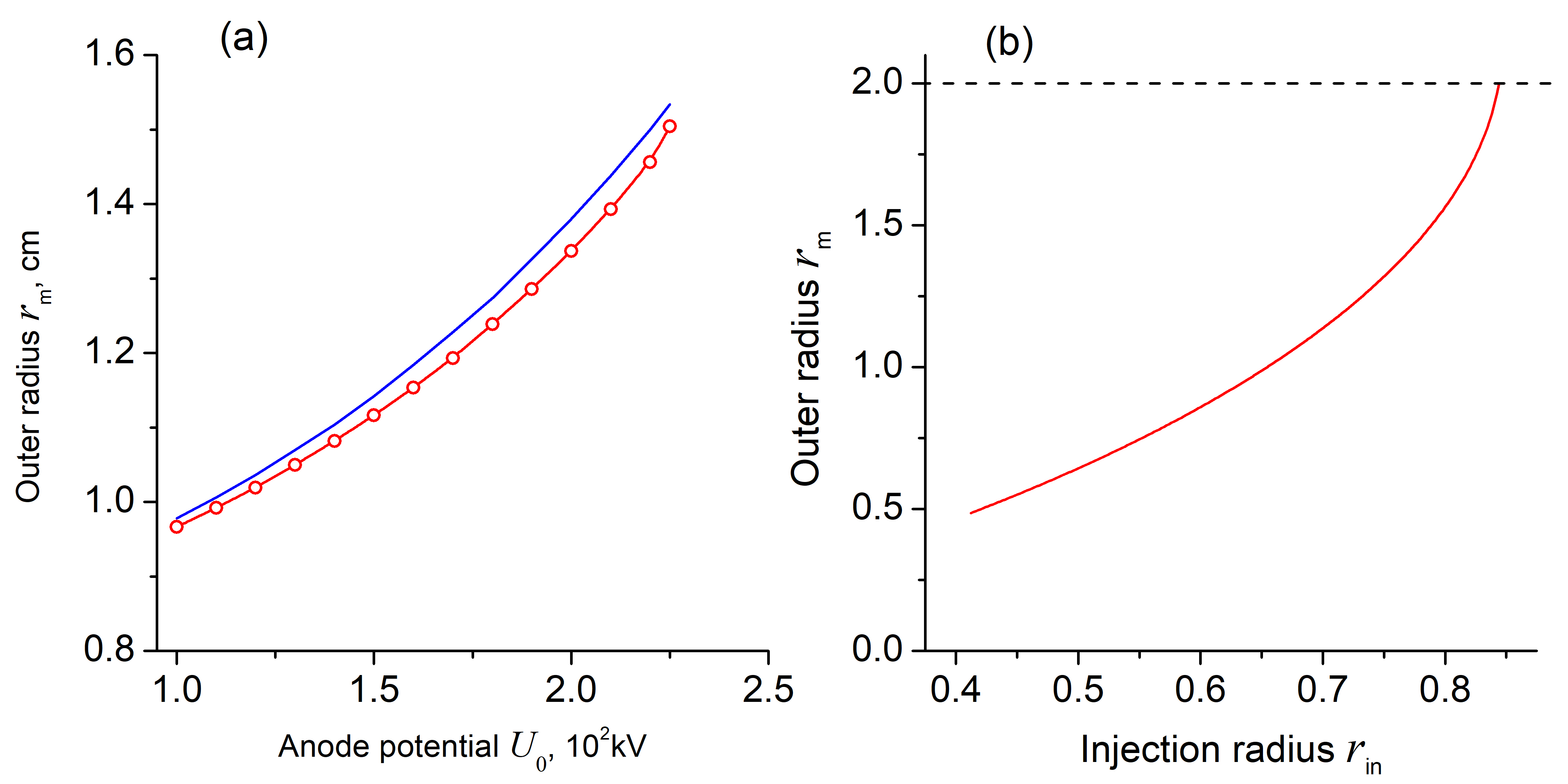}}
	\caption{ (a)  The outer radius $r_m$ of the electron column as a function of the anode potential $U_0$ for $r_a=2$~cm, $r_c=0.25$~cm, $r_{\rm in}=0.8$~cm, $H_0=2$~kG. Solution of Eq.~(\ref{eq19}) (blue solid line) and modeling the column as a set of discrete layers (red dotted line).  (b)  The outer radius of the Brillouin flow as function of the injection radius for $r_c=0.25$~cm, $r_a=2$~cm, $U_0=U_{\rm max}$.}
	\label{fig3}
\end{figure}

As an example, let us consider the system with the same parameters as those used for the calculations in Fig.~\ref{fig2}. The dependence $\rho_m(\Phi_0)$ which follows from Eq.~(\ref{eq19}), expresses itself in Fig.~(\ref{fig3})a in its dimensional form as $r_m(U_0)$.  The equivalent dependence,  $r_m(U_0)$,  obtained by modeling the annular electron column as a set of discrete layers, is also shown in this figure. The permissible anode potential is $U_0\leq 235\,{\rm kV}$.

The outer radius of the electron column is maximal when the dimensionless potential $\Phi_0$ reaches its maximal permissible value $\Phi_{\rm max}$. Equation (\ref{eq19}) with $\Phi_0=\Phi_{max}$ takes the form:
\begin{equation}\label{eq23}
	\rho_{\rm in}^2(\rho_m-\rho_{\rm in})\ln\rho_c+\rho_m^2D(\rho_m,\rho_{\rm in})=0.
\end{equation}
Given the rod radius $\rho_c$, one can find the relation between the inner and outer radii of the electron column, $\rho_{\rm in}$ and  $\rho_m$. Numerical solution of Eq.~(\ref{eq23}) $\rho_m(\rho_{\rm in})$ shown in Fig.~\ref{fig3}b, demonstrates that Brillouin flow with zero anode current at maximal anode potential $\Phi_0=\Phi_{\rm max}$ is not formed, when the injection radius exceeds some critical value, $\rho_{\rm in}>\rho_{\rm crit}$. The value $\rho_{\rm in}=\rho_{\rm crit}$ is defined by taking the root of Eq.~(\ref{eq23}) with $\rho_m=1$. For the system with parameters used in these calculations, $\rho_{\rm crit}=0.422$ ($r_{\rm crit}=0.844$~cm).

\section{Hull cutoff and Buneman-Hartree conditions \label{HC and BH}}

\begin{figure}[htb]
	\centering \scalebox{0.33}{\includegraphics{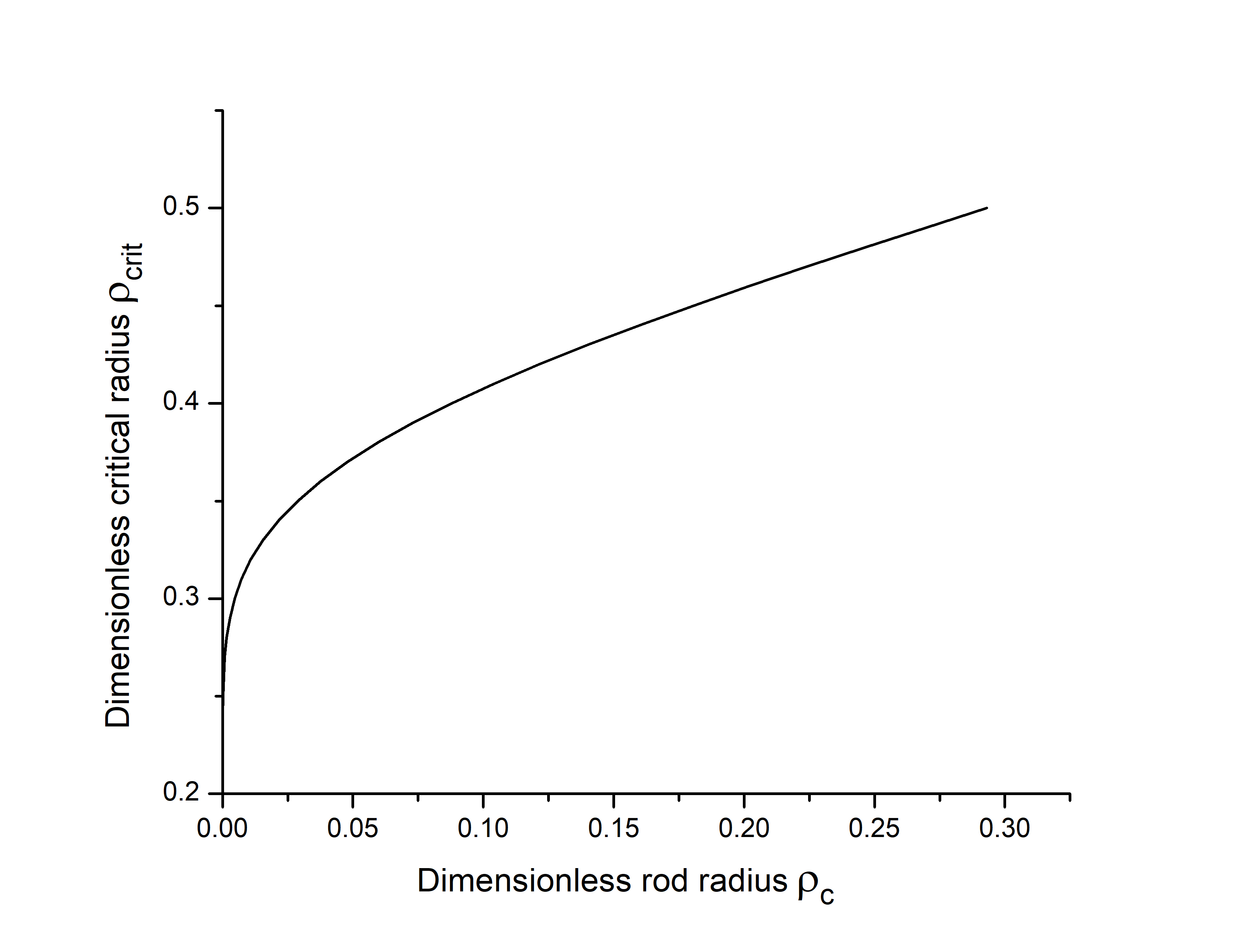}}
	\caption{ The dependence $\rho_{\rm crit}(\rho_c)$. }
	\label{fig4}
\end{figure}

The HC condition determines the maximal value of the dimensionless potential $\Phi_0$ for a given magnetic field $\Omega_H$ which prevents electrons from reaching the anode, i.e., $\rho_m<1$. As shown in Section IV, electrons cannot reach the anode when $\rho_{\rm in}<\rho_{\rm crit}$ for any value of $\Phi_0$ in the permissible interval. The dependence $\rho_{\rm crit}(\rho_c)$,  shown in Fig.~\ref{fig4}, demonstrates, that the critical radius exists when $\rho_c< 0.3$. Otherwise the electrons reach the anode at maximal potential independently  of the value of the injection radius $\rho_{\rm in}$.

However, if $\rho_{\rm in}$ exceeds the critical value $\rho_{\rm crit}$, or the rod radius is large enough, electrons reach anode when the potential $\Phi_0$ exceeds some value $\Phi_{\rm HC}$ smaller than $\Phi_{\rm max}$.  The potential $\Phi_{\rm HC}$ is the solution of Eq.~(\ref{eq19}) for $\rho_m=1$:
\begin{equation}\label{eq25}
\Phi_{\rm HC}\equiv \frac{eU_{\rm BH}}{mc^2}=\frac{\Omega_H^2\ln\rho_c(1-\rho_{\rm in}^4)D(1,\rho_{\rm in})}{4\left[(1-\rho_{\rm in})\ln\rho_c+\rho_{\rm in}^2D(1,\rho_{\rm in})\right]}.
\end{equation}

The conventional BH condition indicates that the rotating electron and the electromagnetic wave, \textit{propagating along the anode surface},  are synchronized, i.e., the wave phase velocity $v_{\rm ph}$ is equal to the electron azimuthal velocity $v_\phi$ . It is assumed that the electron orbit touches the anode. However, if $\rho_{\rm in}<\rho_{\rm crit}$, the outer electron orbit does not touch the anode,  as assumed by the conventional BH condition. For this case it seems reasonable to use an equivalent condition 
\begin{equation}\label{eq27}
	\omega=n\omega_r,
\end{equation}
which is applicable whether the outer electron touches the anode or not. Here $\omega$ is the electromagnetic wave frequency,  $n$ is the azimuthal index of the magnetron electromagnetic eigenmode, and $\omega_r=v_\varphi(r_m)/r_m$ is the electron rotation frequency at the outer surface of the Brillouin flow.

The dimensionless azimuthal velocity $\beta_\varphi=v_\varphi(r_m)/c$ and  rotating frequency $\nu_r$ of electrons at the outer edge $\rho=\rho_m$ of the Brillouin column are equal to the dimensionless azimuthal velocity and rotating frequency of the first injected layer. Using Eqs.~(\ref{eq4}), (\ref{eq6}), and Eq.~(\ref{eq14}), one obtains:
\begin{equation}\label{eq26}
	\beta_\varphi=\nu_r\rho_m=\frac{1}{2}\Omega_H\rho_m\left(1-\frac{\rho_{\rm in}^2}{\rho_m^2}\sqrt{ 1-\Phi_0/\Phi_{\rm max}}\right),
\end{equation}
where $\nu_r$ is the dimensionless electron  rotating frequency. 

\begin{figure}[tbh]
\centering \scalebox{0.5}{\includegraphics{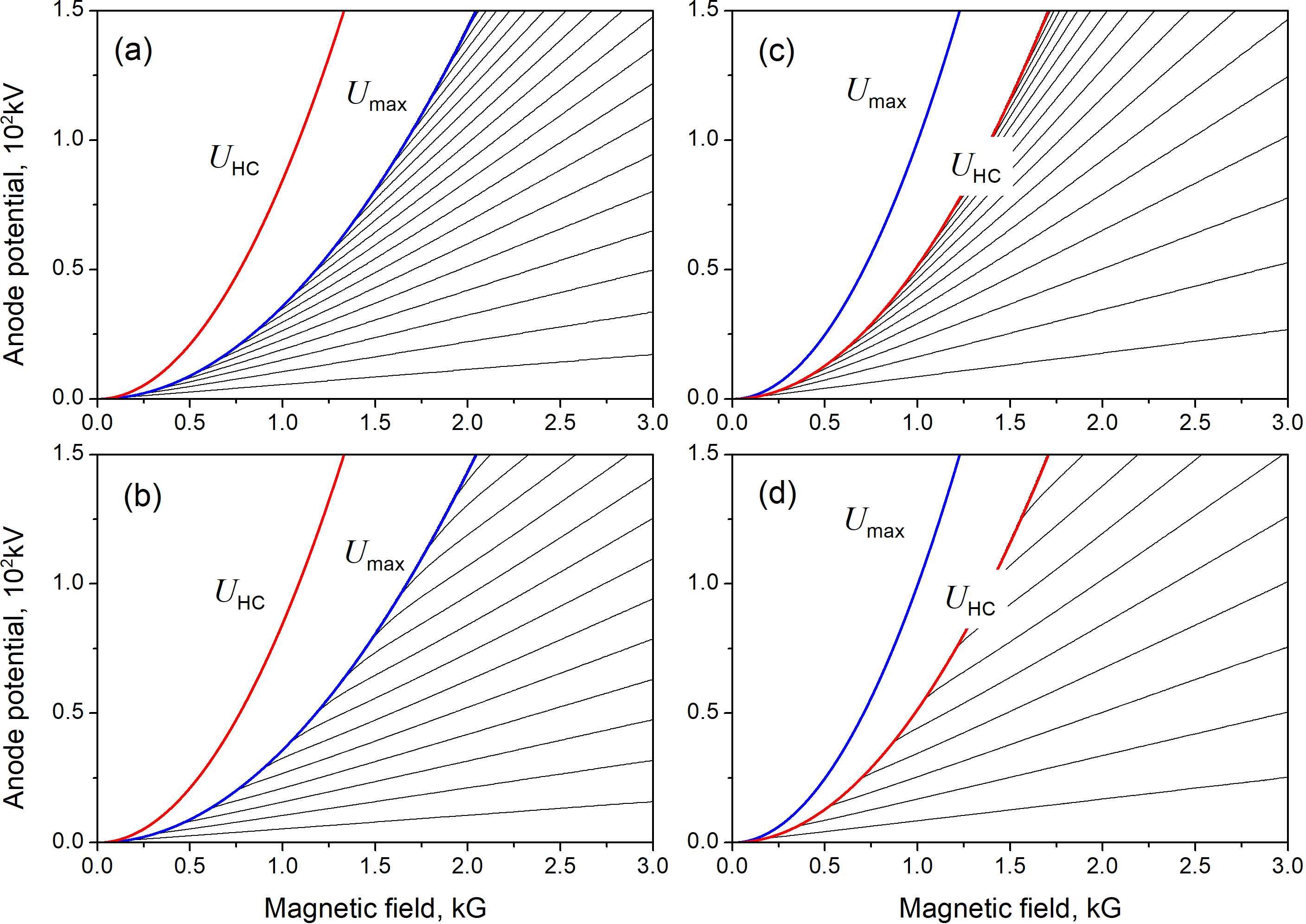}}
\caption{Level lines $\beta\varphi={\rm const}$ [panels (a) and (c)] and $\omega_r={\rm const}$ [panels (b) and (d)] in the plane $(H_0,U_0)$. The maximal permissible potential $U_{\rm max}(H_0)$ and the BH potential $U_{\rm BH}(H_0)$ are shown by solid lines.  (a), (b): $r_{\rm in}=0.6\,{\rm cm}<r_{rm crit}$, Brillouin flow exists for $U_0<U_{\rm max}$.  (c), (d):  $r_{\rm in}=1.0\,{\rm cm}>r_{\rm crit}$, the existence domain is $U_0<U_{\rm HC}$. The anode and the rod radii are $r_a=2.0$~cm and $r_c=0.2$~cm, respectively.}
\label{fig5}
\end{figure}

To demonstrate how the value of SC radius $r_{\rm in}$ affects the BH condition, let us consider two magnetrons with the same anode and rod radii, $r_a=2.0$~cm and $r_c=0.2$~cm, but different the SC radii $r_{\rm in}=0.6\,{\rm cm}<r_{rm crit}$ and $r_{\rm in}=1.0\,{\rm cm}>r_{\rm crit}$. 
The constant $v_\phi(r_m)$ and constant $\omega_r$ curves in the $(H_0, U_0)$ plot are shown in Fig.~\ref{fig5} for both magnetrons. One can see that the system of lines $v_r(r_m)={\rm const}$ (Fig.~\ref{fig5}a,c) or $\omega_r={\rm const}$ (Fig.~\ref{fig5}b,d) are less densely distributed for the magnetron with larger SC radius. It means that BH condition for this magnetron is more sensitive to anode potential $U_0$ variation.

Contrary to magnetrons with conventional cathodes, the Brillouin flow, formed by a SC, is tubular, i.e., there is a region with positive radial gradient of the electron density. Such electron configuration can be unstable to small azimuthal perturbations. This \textit{diocotron} (or slipping stream) instability \cite{Levi-1965, Buneman-1966, SC-Davidson-2001} leads to the development of periodic azimuthal modulation of the electron flow. The diocotron instability in the system studied was observed experimentally and analyzed in \cite{SC-Bliokh-2022} using a quite different model, considering the Brillouin flow as the so-called \textit{squeeze state}\cite{Bliokh-2022a, Dubinov-2020}. The electron rotation frequency $f_r=\omega_r/2\pi$, calculated in the frame of this model, is $f_r\simeq 0.25$~GHz, that agrees well with the experimental data. 

\begin{figure}[tbh]
	\centering \scalebox{0.5}{\includegraphics{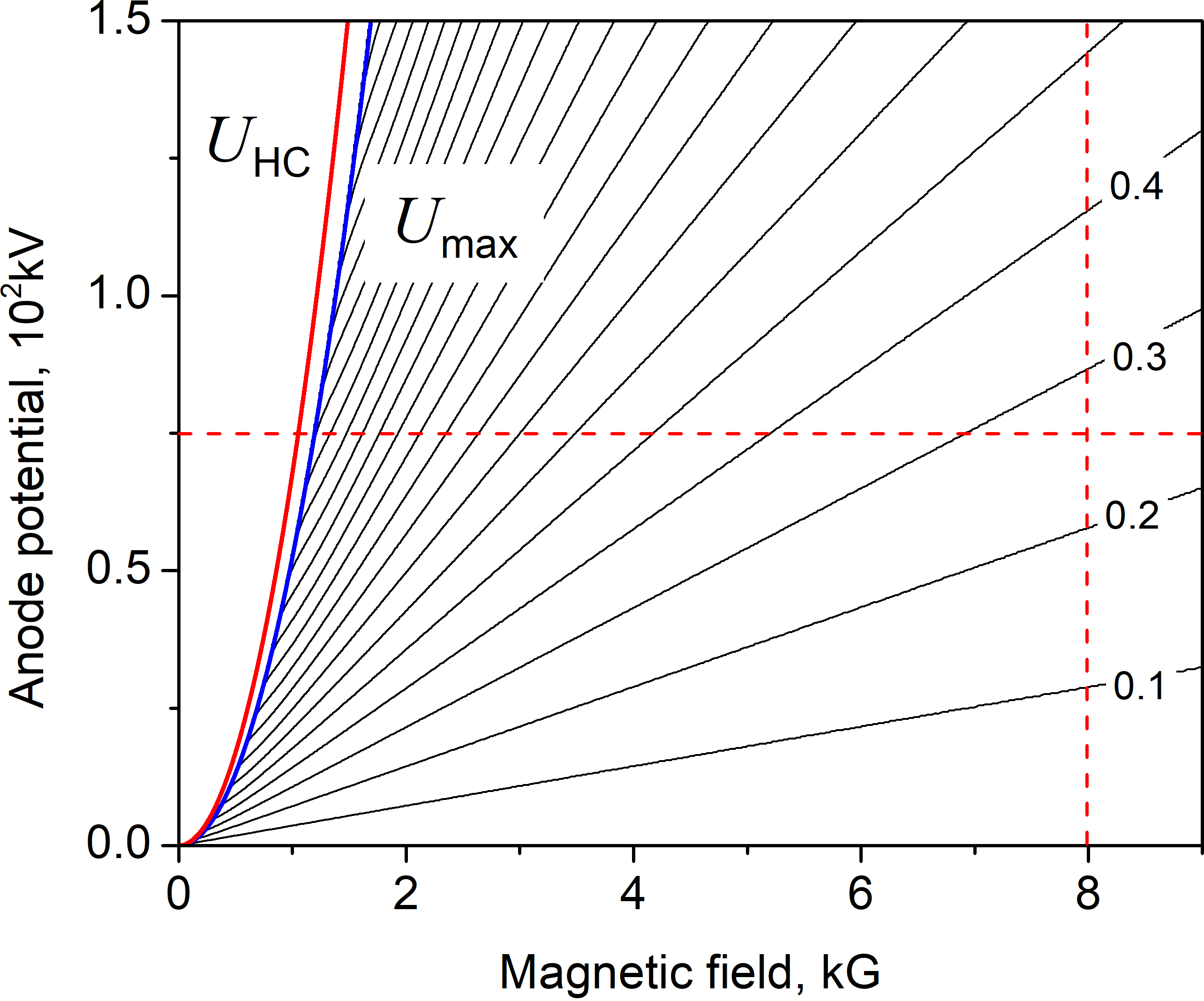}}
	\caption{Level lines $f_r=\omega_r/2\pi={\rm const}$ in the plane $(H_0,U_0)$. The maximal permissible potential $U_{\rm max}(H_0)$ is smaller than the BH potential $U_{\rm BH}(H_0)$.  The anode and the rod radii are $r_a=2.0$~cm and $r_c=0.3$~cm, respectively. The lines $f_r={\rm const}$ are spaced by a 0.1~GHz interval.}
	\label{fig6}
\end{figure}

It seems reasonable to analyze the experimental data, presented in\cite{SC-Bliokh-2022}, using the above developed  model, and compare these two models. The result of this comparison is presented in Fig.~\ref{fig6}. The lines $f_r={\rm const}$ are drawn in an $(H_0,U_0)$ plot with 0.1~GHz intervals. The red dashed lines mark the magnetic field value and anode potential corresponding to values at which the diocotron instability was observed. One can see that the electron rotating frequency is very close to the frequency $f_r\simeq 0.25$~GHz measured and calculated in\cite{SC-Bliokh-2022}.     

\newpage 

\section{discussion}

A physical model for the creation of laminar Brillouin flow in the smooth-bore magnetron fed by a SC was proposed.  It has been shown that the characteristics of the electron flow depends strongly on relations between the radii of the anode $r_a$, the central rod $r_c$, and the emitting cathode $r_{\rm in}$. Laminar flow cannot form when $r_{\rm in}<1.65 r_c$. There is a critical value $r_{\rm crit}$ of the radius $r_{\rm in}$, such that when  $1.65 r_c<r_{\rm in}<r_{\rm crit}$,  the annular laminar flow does not reach neither the anode nor the rod. If the cathode radius exceeds $r_{\rm crit}$, the laminar flow can reach the anode surface. Thus, the choice of the emitter radius allows one to obtain different types of electron flow.     

The tubular Brillouin flow, whose inner and outer surfaces are separated from metal boundaries, can be formed only in magnetrons with a SC. It is well known that this configuration is unstable with respect to the development of the diocotron instability\cite{SC-Davidson-2001}. There is some evidence \cite{SC-Leopold-2023} that the diocotron instability facilitates excitation of electromagnetic oscillations in a cavity magnetron. 

SCs offer advantages over conventional cathodes. Explosive emission plasma forms outside the volume of interest and affects only weakly the magnetron operation because the rod anode distance remains unchanged. The proper choice of  the radius of the emitting region of the SC allows formation of an electron flow with required characteristics.

\begin{acknowledgments}
This work was supported in part by the Technion under Grant 2071448
and in part by the Office of Naval Research Global (ONRG)
under Grant 62909-24-1-2093
\end{acknowledgments}




\end{document}